\documentclass{segabs}

\DeclareMathAlphabet{\mathcal}{OMS}{cmsy}{m}{n}

\usepackage{lipsum}
\usepackage[cmex10]{amsmath}
\usepackage{amsfonts}
\usepackage{bm}
\usepackage{subfigure}
\usepackage{graphicx}

\usepackage{enumitem}

\usepackage{url}

\usepackage{hyperref}

\begin{document}

\onecolumn 

\begin{description}[labelindent=-1.5cm,leftmargin=1cm,style=multiline]

\item[\textbf{Citation}]{Yazeed Alaudah and Ghassan AlRegib (2017) A directional coherence attribute for seismic interpretation. SEG Technical Program Expanded Abstracts 2017: pp. 2102-2106. }

\item[\textbf{DOI}]{\url{ https://doi.org/10.1190/segam2017-17739097.1}}

\item[\textbf{Review}]{Date of publication: August 2017}

\item[\textbf{Data and Codes}]{\url{https://ghassanalregib.com/publications/}}

\item[\textbf{BibTex}] {
\begin{verbatim}
@inbook{doi:10.1190/segam2017-17739097.1,
author = {Yazeed Alaudah and Ghassan AlRegib},
title = {A directional coherence attribute for seismic interpretation},
booktitle = {SEG Technical Program Expanded Abstracts 2017},
chapter = {},
pages = {2102-2106},
year = {2017},
doi = {10.1190/segam2017-17739097.1},
URL = {https://library.seg.org/doi/abs/10.1190/segam2017-17739097.1},
eprint = {https://library.seg.org/doi/pdf/10.1190/segam2017-17739097.1}
}
\end{verbatim}
}

\item[\textbf{Contact}]{\href{mailto:alaudah@gatech.edu}{alaudah@gatech.edu}  OR \href{mailto:alregib@gatech.edu}{alregib@gatech.edu}\\ \url{http://ghassanalregib.com/} }
\end{description}

\thispagestyle{empty}
\newpage
\clearpage
\setcounter{page}{1}

\title{A Directional Coherence Attribute for Seismic Interpretation}

\author{Yazeed Alaudah and Ghassan AlRegib \\ Center for Energy and Geo Processing (CeGP) at Georgia Tech and KFUPM}

\lefthead{Alaudah \& AlRegib}

\maketitle
\righthead{A Directional Coherence Attribute for Seismic Interpretation}

\begin{abstract}

The coherence attribute is one of the most commonly used attributes in seismic interpretation. In this paper, we propose building on the recently introduced Generalized Tensor-based Coherence (GTC) attribute to make it directionally selective. This directional selectivity is achieved by selecting a directional Gaussian preprocessing kernel and applying a 3D rotational matrix to its covariance matrix. By weighing traces, or voxels, in the analysis cube by their relative proximity to the reference trace or voxel, this approach greatly enhances the clarity of the attribute. Furthermore, by making these weights directional,  the proposed attribute gives interpreters greater freedom in exploring and understanding the seismic data. Various results from the Netherlands North Sea F3 block show that this approach greatly enhances the clarity of the coherence attribute and can highlight structures that are not visible using the traditional C3, or GTC coherence. 

\end{abstract}

\section{Introduction}

The coherence attribute has proven to be a very useful attribute for highlighting structural and stratigraphic discontinuities such as faults, fractures, and channels in 3D seismic volumes. The coherence attribute was first proposed by \cite{Bahorich1995}, and it was based on the normalized cross-correlation between each trace and its adjacent traces. \cite{Marfurt2000} then proposed a multi-trace semblance-based coherence algorithm that was more robust to noise and improved vertical resolution. Later, \cite{Gersztenkorn1999}  proposed an improved coherence algorithm, called C3 coherence, which is based on the eigenstructure of covariance matrices of windowed seismic traces. Recently, there has been renewed interest in the coherence attribute.  \cite{Yang2015a} proposed a computationally efficient coherence algorithm based on a normalized information divergence criterion that avoids directly calculating the eigenvalues of the covariance matrix. In addition, \cite{Li2014a} combined spectral decomposition and complex coherence computation to map discontinuities at different scales. Finally, to avoid false low-coherence values in steeply dipping structures, \cite{Sui2015a} proposed a coherence algorithm that analyzes the eigenstructure of the \textit{spectral} amplitudes of seismic traces.

The C3 coherence is based on the eigenstructure of the covariance matrix of the zero-mean traces in the analysis cube. In our recent work \cite[]{yazeedEAGE2016}, we have shown that this is analogous to unfolding a $3^{rd}$-order analysis tensor in a single mode and computing the covariance matrix of that mode. By unfolding the tensor along the two other modes, and repeating the process, then assigning each coherence attribute from each mode a different color we can significantly enhance the amount of detail that the C3 coherence can extract from seismic volumes. Using this insight, we proposed the Generalized Tensor-Based Coherence (GTC) attribute in \cite[]{yazeedEAGE2016}. The GTC attribute can be viewed as a generalization of the C3 coherence attribute that was proposed by \cite{Gersztenkorn1999}.

In this paper, we further expand on the GTC attribute we proposed earlier by enhancing its directional selectivity using a directional Gaussian preprocessing kernel, rotated to arbitrary angles using 3D rotational matrices. This enables this enhanced directional coherence attribute to be to able to highlight faults, fractures, channels, and other subsurface structures characterized by their high directionality. We show that this directional selectivity gives interpreters much more flexibility with exploring post-migration seismic data than traditional coherence attributes.




The structure of this paper is as follows: First, we introduce the GTC attribute. Then, we introduce the multivariate Gaussian kernel that is used as a preprocessing step for the GTC attribute. We then show how to make the GTC attribute directionally selective. Finally, before we conclude the paper, we show several results comparing the C3 coherence with the GTC attribute and our proposed directional coherence attribute. 

\section{Generalized Tensor-Based Coherence (GTC) Algorithm}

Given a migrated 3D seismic volume, the coherence attribute for each voxel in the volume is computed within a small 3D analysis cube of size $I_1 \times I_2 \times I_3$. The subscripts $1$,$2$ and $3$ throughout this paper refer to the dimensions along time (or depth), inline, and crossline respectively. Each analysis cube can be represented as a $3^{rd}$ order tensor $ \mathcal{A} \in \mathbb{R}^{I_1 \times I_2 \times I_3}$ that we refer to as the analysis tensor. To compute the covariance matrices of  this tensor, we unfold the tensor along its three modes. In general, mode-$n$ unfolding of an $N$-th order tensor results in a matrix $\mathbf{A}_{(n)}$ of size $I_n$ by $(I_1 \cdots  I_{n-1} I_{n+1}  \cdots  I_N)$  where the tensor element indexed by $(i_1,i_2, \cdots, i_N)$  now corresponds to the element $(i_n,j)$ in  $\mathbf{A}_{(n)}$ where
\begin{equation}
j = 1+ \sum_{\substack{k=1 \\ k\neq n}}^N (i_k -1) \prod_{\substack{m=1 \\ m\neq n}}^{k-1} I_m.
\end{equation}

For additional details, see \cite{Santiago2005}. Thus unfolding the tensor along its three modes results in three matrices: the $I_1 \times I_2 I_3$ mode-1 matrix  $\mathbf{A}_{(1)}$ unfolded along the time (depth) dimension, the $I_2 \times I_1 I_3$ mode-2 matrix  $\mathbf{A}_{(2)}$ unfolded along the inline dimension, and the $I_3 \times I_1 I_2$ mode-3 matrix  $\mathbf{A}_{(3)}$ unfolded along the crossline dimension. The covariance matrices are then given by

\begin{equation}
\mathbf{C}_1 = (\mathbf{A}_{(1)} - \underset{I_1 \times 1}{\mathbf{1}} \times \bm{\mu}_1)^T(\mathbf{A}_{(1)} - \underset{I_1 \times 1}{\mathbf{1}} \times \bm{\mu}_1),
\end{equation}  


where $\bm{\mu}_1$ is a row vector of length $I_2I_3$ containing the means of all columns of $\mathbf{A}_{(1)}$, and $ \underset{I_1 \times 1}{\mathbf{1}}$ is a column vector of ones of length $I_1$. $\mathbf{C}_2$ and $\mathbf{C}_3$ are also computed in a similar fashion. These covariance matrices are positive semi-definite matrices and  thus all their eigenvalues are non-negative. If we denote the ranked eigenvalues of $\mathbf{C_1}$ as $\bm{\lambda}^{(1)}= \{\lambda_1^{(1)}, \lambda_2^{(1)}, \cdots  , \lambda_{I_2I_3}^{(1)}\}$, and similarly for $\mathbf{C}_2$ and $\mathbf{C}_3$, then the coherence attributes of the three different modes are given as the ratios of the largest eigenvalue of the covariance matrix to its trace. Specifically,
\begin{equation}
E_c^{(1)} = \frac{\lambda_1^{(1)}}{Tr(\mathbf{C_1})}  =  \frac{\lambda_1^{(1)}}{\sum_{i=1}^{I_2I_3} \lambda_i^{(1)} }, 
\end{equation}
\begin{equation}
E_c^{(2)} = \frac{\lambda_1^{(2)}}{Tr(\mathbf{C_2})}  =  \frac{\lambda_1^{(2)}}{\sum_{i=1}^{I_1I_3} \lambda_i^{(2)} }, 
\end{equation}
\begin{equation}
\mathrm{and} ~ E_c^{(3)} = \frac{\lambda_1^{(3)}}{Tr(\mathbf{C_3})}  =  \frac{\lambda_1^{(3)}}{\sum_{i=1}^{I_1I_2} \lambda_i^{(3)} }.
\end{equation}

Here, $E_c^{(1)}$ corresponds to the C3 coherence attribute that was proposed by \cite{Gersztenkorn1999}. By combining the C3 attribute ($E_c^{(1)}$)  with the coherence estimates of the analysis tensor unfolded along mode-2 ($E_c^{(2)}$) and mode-3 ($E_c^{(3)}$) in different color channels, we arrive at the GTC coherence attribute that was proposed by \cite{yazeedEAGE2016}. 


\section{The Preprocessing Gaussian kernel}
Most coherence algorithms treat all traces in an analysis tensor equally regardless of their proximity to the reference trace. This is understandable if the analysis tensor dimensions were very small. However, as the dimensions of the tensor become larger, this introduces noise to the different covariance matrices. This is the case when either or all the dimensions $I_1$, $I_2$, and $I_3$   are greater than or equal to 5. 

\cite{yazeedEAGE2016} proposed a preprocessing step that weighs different traces by weights relative to their proximity to the reference trace and thus eliminates this problem. Given the 3D analysis tensor $\mathcal{A}$, we can preprocess it by taking its element-wise product with a 3-dimensional Gaussian kernel of the same size as $\mathcal{A}$. We can write the preprocessed analysis tensor as
\begin{equation}
\widetilde{\mathcal{A}} = \mathcal{A} \odot \mathcal{G},
\end{equation} 
where $\odot$ is the element-wise product, and $\mathcal{G}$ is the multivariate Gaussian kernel given by

\begin{equation}\label{gauss}
\mathcal{G}(\mathbf{x}; \bm{\mu}, \Sigma) =  e^{-\frac{1}{2}(\mathbf{x} -\bm{\mu})^T \Sigma^{-1}(\mathbf{x} -\bm{\mu})}.
\end{equation} 
Here, $\mathbf{x} = \{z,x,y\}$ represents the voxels in the seismic volume,  $\bm{\mu} = \{z_0,x_0,y_0\}$ is the reference voxel, and $\Sigma$ is the $3\times3$ covariance matrix of $\mathcal{G}$. The expression in equation (\ref{gauss}) is equivalent to a multivariate Gaussian \emph{distribution} multiplied by $(2\pi)^{\frac{3}{2}} | \Sigma^{\frac{1}{2}}|$. The values of $\Sigma$ describe the shape of the multivariate Gaussian kernel. The interpreter can select these values to give more emphasis to the coherence attribute extracted along any unfolding mode, or any combination of unfolding modes. Extracting the GTC  coherence attribute from $\widetilde{\mathcal{A}}$ as opposed to $\mathcal{A}$ greatly enhances the results.


\section{Making the GTC directionally selective}

\begin{figure}
    \centering
    \begin{subfigure}{}
        \centering
        \includegraphics[width=0.40\textwidth]{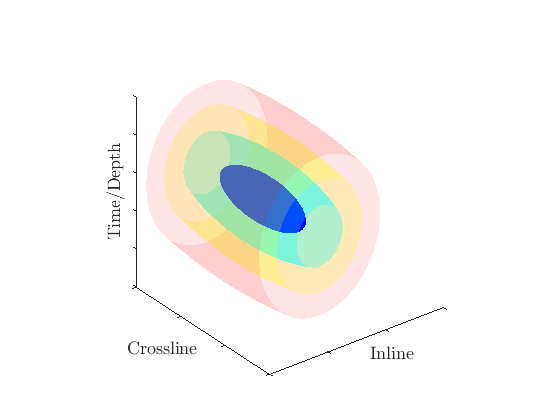}
    \end{subfigure}%
    \begin{subfigure}{}
        \centering
        \includegraphics[width=0.40\textwidth]{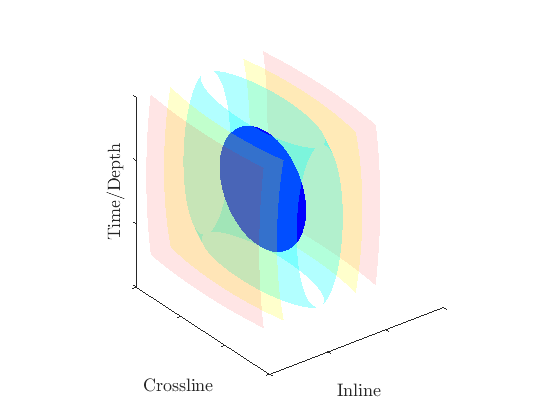}
    \end{subfigure}
    \caption{Examples of a 3D directional Gaussian preprocessing kernel. The blue, cyan, yellow, and red contours refer to values corresponding to $0.5\sigma, \sigma, 1.5 \sigma$ and $2\sigma$ of the Gaussian kernel respectively.}
    \label{fig:kernel}
\end{figure}

By selecting the values of $\Sigma$ in such a way that the Gaussian kernel is not symmetric along all directions, we obtain a directionally selective kernel similar to those in figure \ref{fig:kernel}.   

To rotate this kernel along different directions and angles, we replace the covariance matrix $\Sigma$ in equation \ref{gauss} by its rotated version $\Sigma_\theta$, where:
\begin{equation}
    \Sigma_\theta = R \Sigma R^T.
\end{equation}

Here, $R$ is an orthogonal 3D rotational matrix. This can be either rotated along the time direction using 
\begin{equation} \label{eq:r1}
    R_1 =  
    \begin{pmatrix}
    1 & 0 & 0 \\
    0 & \cos \theta & -\sin \theta \\
    0 & \sin \theta & \cos \theta 
    \end{pmatrix},
\end{equation}
the inline direction using  
\begin{equation} \label{eq:r2}
    R_2 =  
    \begin{pmatrix}
    \cos\theta & 0 & \sin \theta\\
    0 & 1 & 0 \\
    -\sin\theta & 0 & \cos\theta
    \end{pmatrix},
\end{equation}
or the crossline direction using
\begin{equation} \label{eq:r3}
    R_3 =  
    \begin{pmatrix}
    \cos \theta & -\sin \theta & 0\\
    \sin \theta & \cos \theta & 0\\
    0 & 0 & 1
    \end{pmatrix}.
\end{equation}
 
In equations \ref{eq:r1}, \ref{eq:r2}, and \ref{eq:r3}, $\theta$ refers to the angle of the rotation of the kernel.

\section{Results}

\begin{figure}[ht]
  \centering
  \subfigure[Seismic amplitude]{\includegraphics[width=0.2\textwidth]{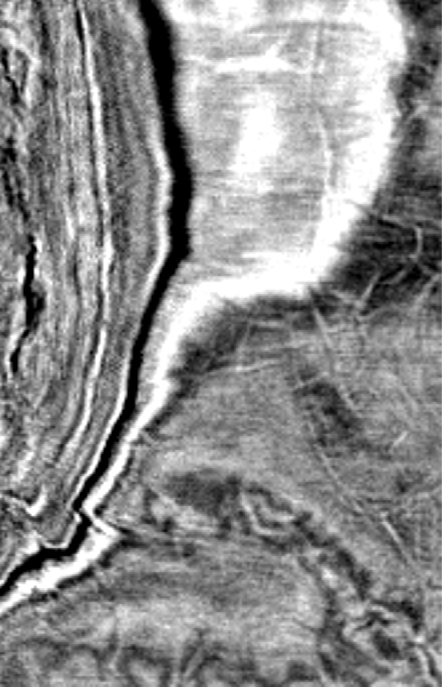}\label{fig:2a}}
  \subfigure[C3 coherence]{\includegraphics[width=0.1948\textwidth]{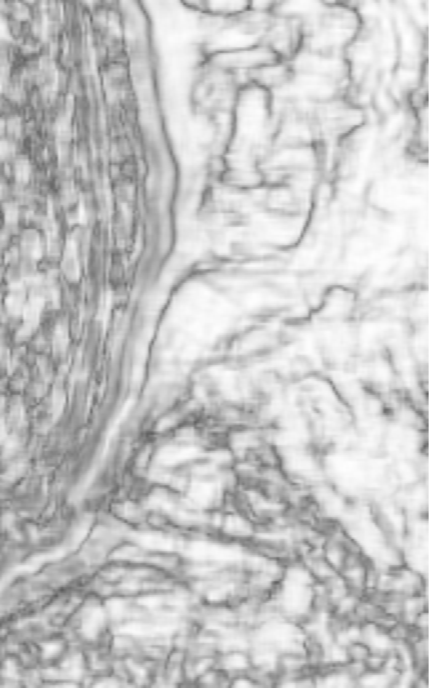}\label{fig:2b}}
  \subfigure[GTC coherence in \cite{yazeedEAGE2016}]{\includegraphics[width=0.195\textwidth]{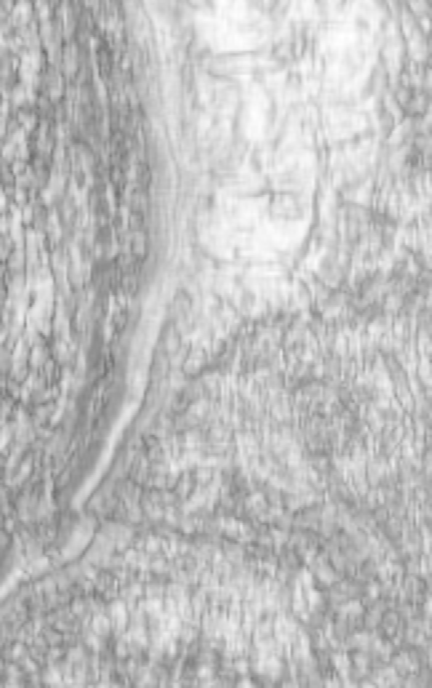}\label{fig:2c}}
  \subfigure[Proposed directional coherence with $\theta = 40 ^\circ$ ]{\includegraphics[width=0.2\textwidth]{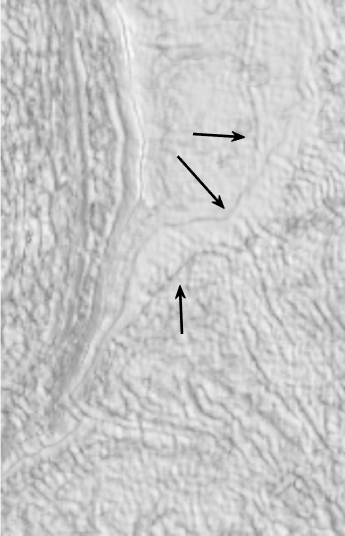}\label{fig:2d}}
  \subfigure[Proposed directional coherence with $\theta = 75 ^{\circ}$]{\includegraphics[width=0.2\textwidth]{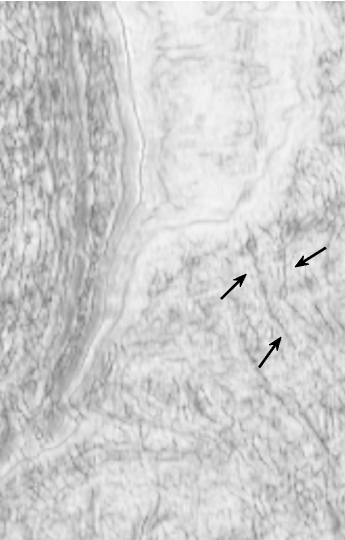}\label{fig:2e}}
  \subfigure[Proposed directional coherence with $\theta = 150 ^{\circ}$]{\includegraphics[width=0.2\textwidth]{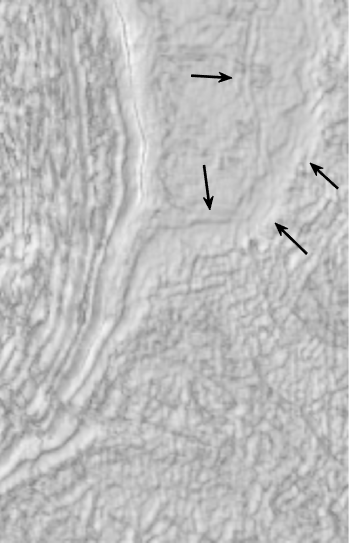}\label{fig:2f}}

  \caption{Coherence values for time section 1600ms in the Netherlands North Sea F3 block. Sub-figures \ref{fig:2d}, \ref{fig:2e}, and \ref{fig:2f} show our proposed directional coherence using different angles $\theta$. Black arrows indicate the boundaries of a channel formation not visible in the seismic amplitude \ref{fig:2a}, the C3 coherence \ref{fig:2b}, or the GTC coherence \ref{fig:2c}.}
  \label{fig:angles}
\end{figure}

To demonstrate the effectiveness of our approach, we apply the proposed directional attribute on the 1600ms time section of the Netherlands offshore F3 block in the North Sea provided by \cite{F3_data}. We also apply the GTC and C3 coherence attributes for comparison. Figure \ref{fig:2a} and \ref{fig:2b} shows the seismic amplitude of part of the time section, along with the computed C3 coherence. Figure \ref{fig:2c} shows the computed GTC coherence, while figures \ref{fig:2d},  \ref{fig:2e}, and  \ref{fig:2f} shows the proposed directional coherence for three different angles, all rotated along the time direction. It is easy to observe the higher level of detail showed by the directional coherence compared to the C3 or GTC coherence. Black arrows in figure \ref{fig:angles} show a channel formation that wasn't visible using the other coherence attributes. 

Figure 3, 4, 5 and 6 shows the seismic amplitude, the C3 coherence, the GTC coherence, and the proposed directional coherence, respectively for the entire 1600ms time section. Figure \ref{fig:7} shows these attributes side by side in grayscale. It is easy to see that the GTC attribute shows much more detail than the C3. Also, the proposed directional attribute highlights directional features such as faults and fractures much more clearly than the GTC attribute. This can be observed by comparing figure \ref{fig:7c} to figures \ref{fig:7a} and \ref{fig:7b}.


\newsavebox{\smlmatone}
\savebox{\smlmatone}{$\Sigma = \left(\begin{smallmatrix} 2 & 0 & 0 \\
    0 & 2 & 0\\ 0 & 0 & 2  \end{smallmatrix}\right)$}
    
\newsavebox{\smlmattwo}
\savebox{\smlmattwo}{$\Sigma = \left(\begin{smallmatrix} 5 & 0 & 0 \\
    0 & 5 & 0\\
    0 & 0 & 1.5 \end{smallmatrix}\right)$}

\begin{figure*}[hb] 
  \begin{minipage}[]{0.5\linewidth}
    \centering
    \includegraphics[width=0.99\linewidth]{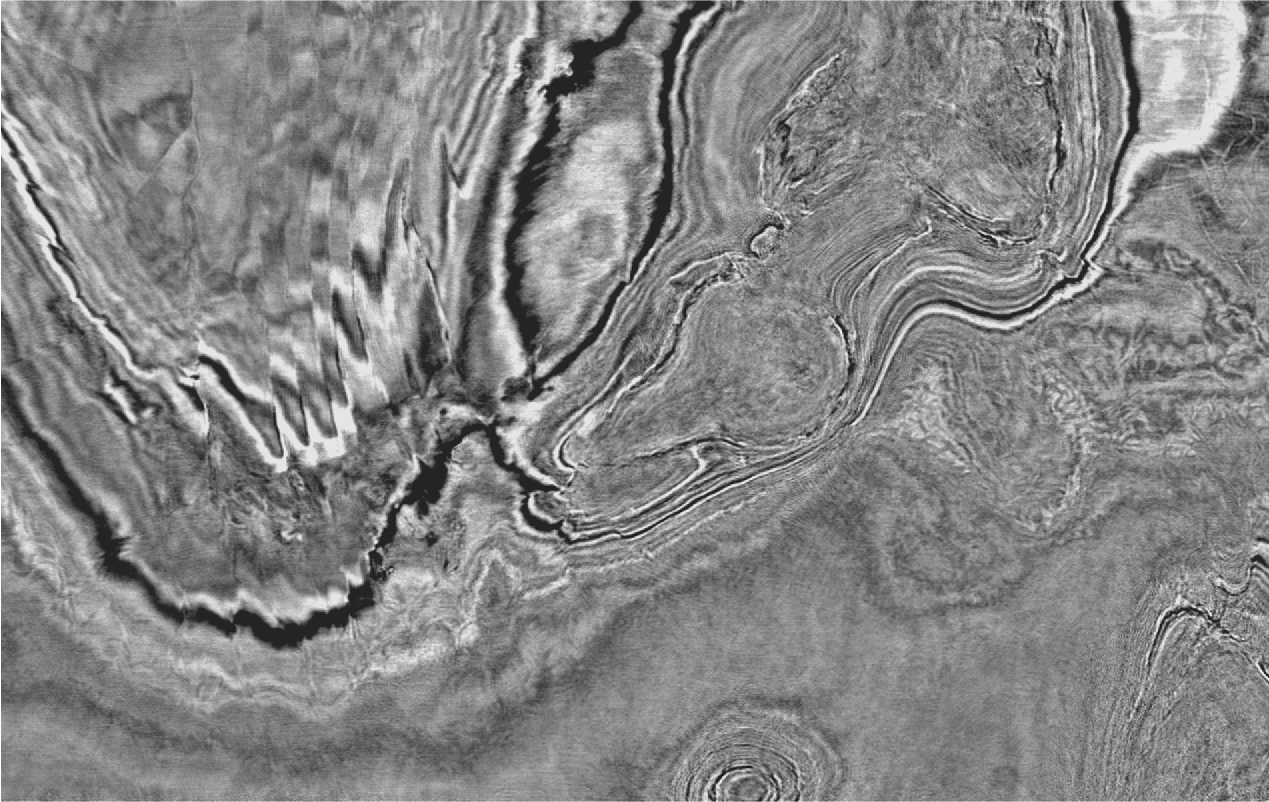} 
    \caption{1600ms time section of the Netherlands F3 block} 
    \vspace{4pt}
  \end{minipage}
  \begin{minipage}[]{0.5\linewidth}
    \centering
    \includegraphics[width=0.99\linewidth]{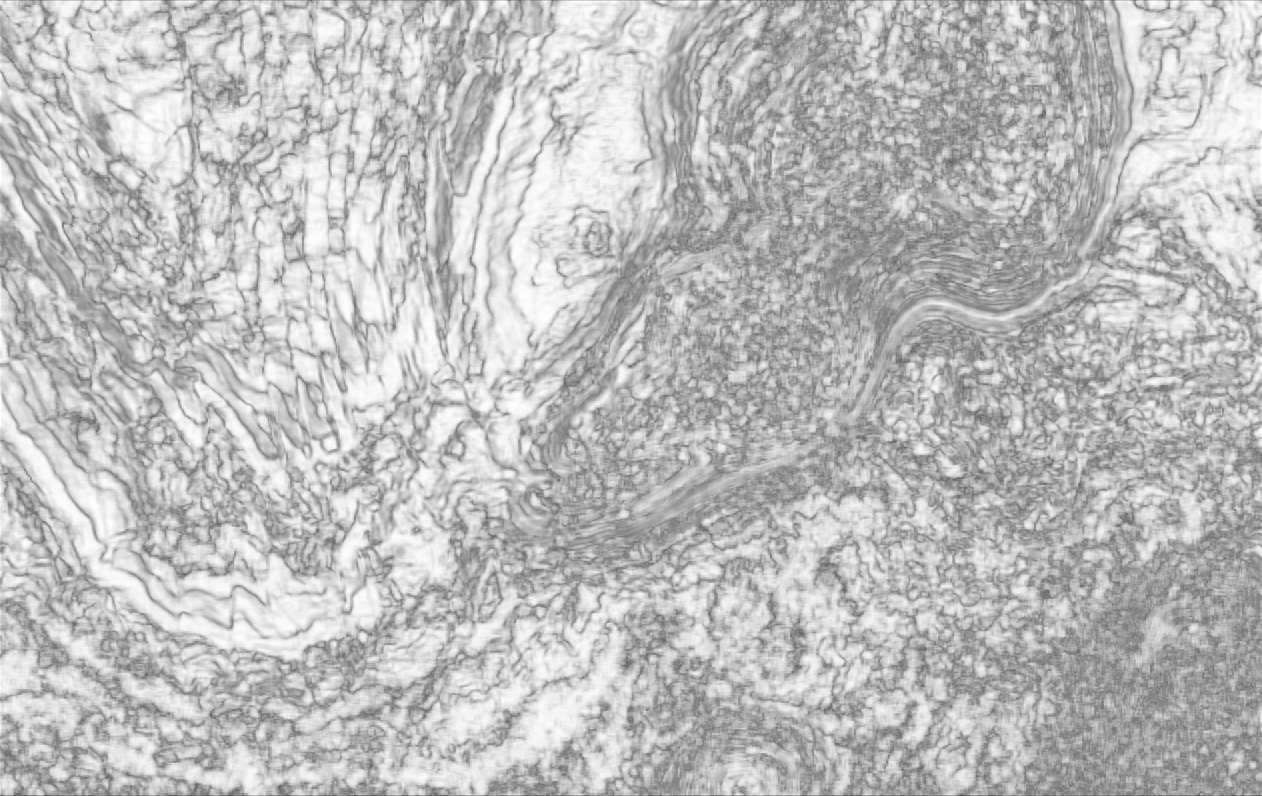} 
    \caption{The C3 coherence attribute} 
    \vspace{4pt}
  \end{minipage} 
  \begin{minipage}[]{0.5\linewidth}
    \centering
    \includegraphics[width=0.99\linewidth]{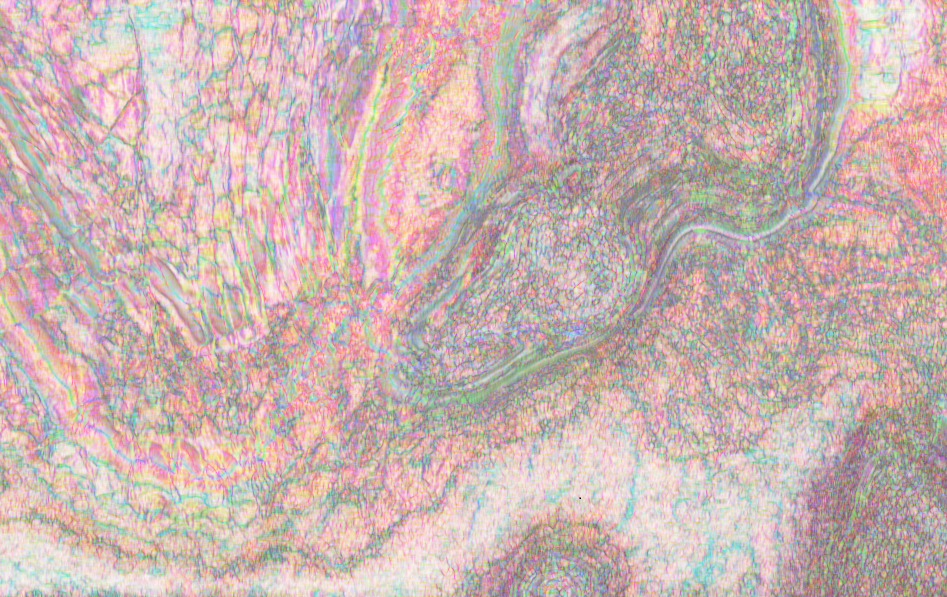} 
    \caption{The GTC coherence attribute with: ~\usebox{\smlmatone} and $theta = 160^\circ$}
    \vspace{4pt}
  \end{minipage}
  \begin{minipage}[]{0.5\linewidth}\label{fig:6}
    \centering
    \includegraphics[width=0.99\linewidth]{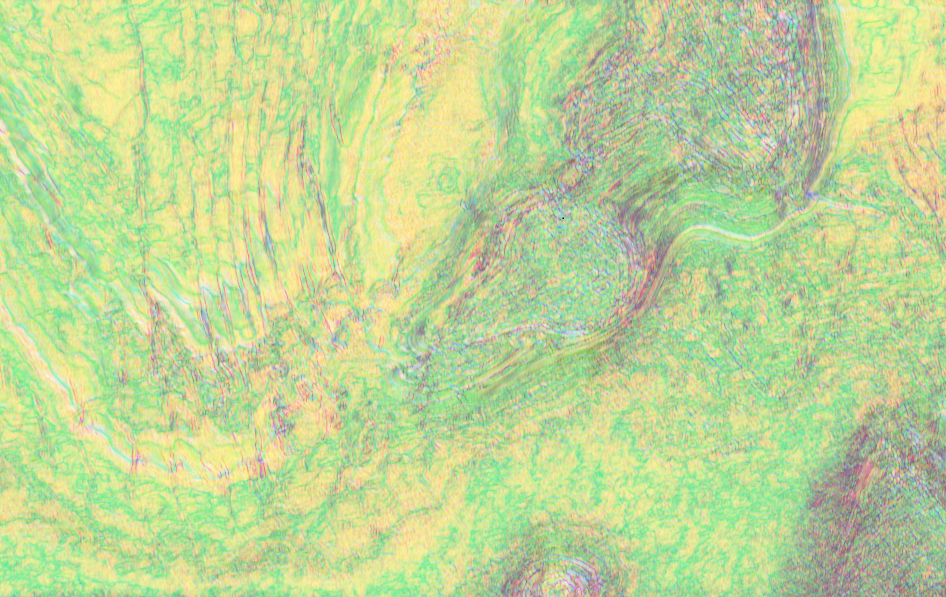} 
    \caption{The proposed directional GTC coherence attribute with ~\usebox{\smlmatone}
    , $\theta = 160^\circ$ ,and rotated along the time axis.} 
    \vspace{4pt}
    \end{minipage} 
\end{figure*}

\begin{figure*}[h]
  \centering
  \subfigure[C3 coherence]{\includegraphics[width=0.33\textwidth]{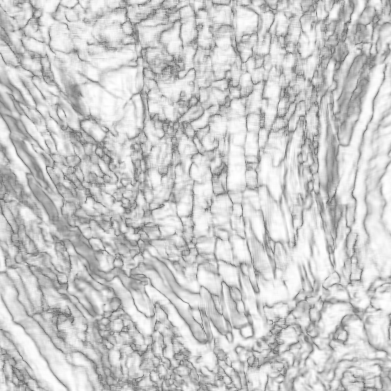}\label{fig:7a}}
  \subfigure[GTC coherence in grayscale]{\includegraphics[width=0.33\textwidth]{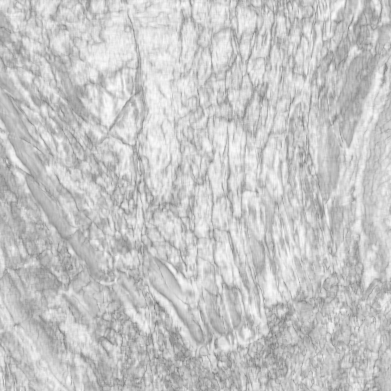}\label{fig:7b}}
  \subfigure[Proposed directional coherence in grayscale ($\theta = 160^\circ$)]{\includegraphics[width=0.33\textwidth]{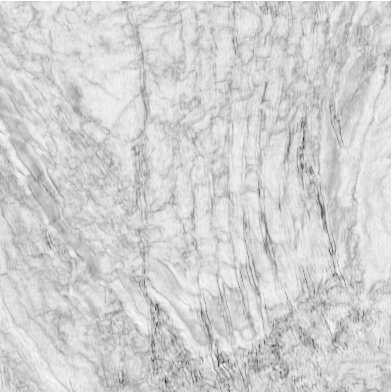}\label{fig:7c}}

  \caption{The C3 coherence along with grayscale versions of the GTC coherence and the proposed directional coherence (using the same values as in figure 6). All three use $5\times5 \times
  5$ analysis tensors. Note the various faults and fractures along the vertical direction in \ref{fig:7c} are more clearly visible compared to \ref{fig:7a} or \ref{fig:7b}.}
  \label{fig:7}
\end{figure*}

\section{Conclusions}

In conclusion, we have proposed a new directional coherence attribute that is an extension of the generalized tensor-based coherence attribute (GTC). We have shown that the proposed attribute can highlight subtle subsurface features such as channels, faults, and fractures that were not visible in the C3 or GTC coherence attributes. Various results from the Netherlands North Sea F3 block show the effectiveness of this attribute compared to the other methods. 

\section{Acknowledgments}

The authors would like to acknowledge the support of the Center for Energy and Geo Processing (CeGP) at the Georgia Institute of Technology and King Fahd University of Petroleum and Minerals (KFUPM).

\twocolumn

\bibliographystyle{seg}  
\bibliography{main.bib}

\end{document}